# An Empirical Evaluation of Ensemble Adjustment Methods for Analogy-Based Effort Estimation

Mohammad Azzeh[1], Ali Bou Nassif[2], Leandro Minku[3]

[1]*Department of Software Engineering, Applied Science University, Jordan. E-mail: m.y.azzeh@asu.edu.jo*
[2]*Department of Electrical and Computer Engineering, University of Western Ontario, Canada. E-mail: abounas@uwo.ca*
[3]*School of Computer Science, University of Birmingham, UK. E-mail: L.L.Minku@cs.bham.ac.uk*

**Abstract**—
***Context***: Effort adjustment is an essential part of analogy-based effort estimation, used to tune and adapt nearest analogies in order to produce more accurate estimations. Currently, there are plenty of adjustment methods proposed in literature, but there is no consensus on which method produces more accurate estimates and under which settings.
***Objective***: This paper investigates the potential of ensemble learning for variants of adjustment methods used in Analogy-Based Effort Estimation. The number *k* of analogies to be used is also investigated.
***Method***: We perform a large scale comparison study where many ensembles constructed from *n* out of 40 possible valid variants of adjustment methods are applied to 8 datasets. The performance of each method was evaluated based on standardized accuracy and effect size.
***Results***: The results have been subjected to statistical significance testing, and show reasonable significant improvements on the predictive performance where ensemble methods are applied.
***Conclusion***: Our conclusions suggest that ensembles of adjustment methods can work well and achieve good performance, even though they are not always superior to single methods. We also recommend constructing ensembles from only linear adjustment methods, as they have shown better performance and were frequently ranked higher.

## 1 INTRODUCTION

Analogy-Based Effort Estimation (EBA) is a commonly used method for predicting the most likely software development effort [1][2]. It is based on the assumption that software projects with similar characteristics have similar effort values [18][22][37][49]. Reusing efforts of the selected analogies directly without considering revision is less accurate [4][19]. Therefore, an adjustment technique should be applied to calibrate and tune the generated estimate based on the characteristics of both source and target projects. The goal of using adjustment is to minimize differences between a new project and its nearest analogies, and therefore increase EBA's accuracy.

Many adjustment methods have been proposed in the past twenty years [4], but as of yet, there is no univocal conclusion as to which adjustment method integrated with EBA produces the most accurate predictions, and under which settings. However, Azzeh [4]'s replication study reported an important insight. He showed that, even though no particular method is significantly superior to others, guidelines can be given to explain how and under what conditions to use each of the existing methods. It has been concluded that each method favors: 1) different feature set, 2) different number of nearest analogies (*k*) and 3) specific type of features (i.e. continuous or categorical). Moreover, the results from that study showed that some adjustment methods cannot outperform conventional EBA over some datasets. For these reasons it was difficult to recommend a particular method against others over a particular dataset. We believe that it would be more promising to combine existing methods in order to benefit from their individual advantages (and consequently improve the accuracy of adjusted EBA) rather than to create a new adjustment method.

The literature on predictive methods for software effort estimation has shown that combining several predictive models into an ensemble can produce more accurate results than single models [44]. Prior work on ensemble methods in the area of data mining also reports that ensembles can produce accurate results in comparison to single models, if not superior [46][47][48]. The idea behind the success of ensembles is that the accurate predictions given by some of its models to a given example can patch the mistakes given by others to this example [44]. In this way, the overall accuracy of the ensemble can be better than the individual accuracies of its base models. In order to achieve that, it is well accepted that the base models composing the ensemble should be diverse, i.e., they should make different mistakes on the same data points [56][60]. If they make the same mistakes, then the ensemble will also make the same mistakes as the individual models, and its performance will be no better than the individual performances. In other words, ensembles of non-diverse models are unsuccessful in improving the accuracy of these models.

Even though ensembles of software effort estimation models have been increasingly studied in software engineering, this is the first study that attempts to combine adjustment methods into ensembles. It is not known whether ensembles of adjustment methods would be successful in improving the accuracy of the calibration of EBA, and consequently the accuracy of EBA itself. In particular, it is not known whether different adjustment techniques behave diversely enough, i.e., if their amount of diversity is enough to lead to improvements in performance. If they do not, then combining these different techniques into an ensemble may not really improve performance. The main objective of this study is thus to investigate the potential of ensembles of adjustment methods for EBA.

With that in mind, this study aims at answering the following research questions:

RQ1. Is there evidence that ensembles improve the accuracy of adjusted EBA?

RQ2. Which approach is better for adjustment, Linear of Non-Linear methods?

RQ3. Is there evidence that using different $k$ analogies makes adjustment methods behave diversely?

The main contributions of this paper are the following:

1) An evaluation of each adjusted EBA variant over all datasets to identify the ones that are actual prediction methods based on *Standardized Accuracy (SA)* measure and effect size.
2) Ranking and clustering of actual prediction methods using Scott-Knott to identify the best methods with smallest *Mean Absolute Error*.
3) A new approach to build ensembles of adjustment methods based on Scott-Knott test method and Borda count procedure. This method can work well when all best methods identified by Scott-Knott are statistically similar. Existing methods such as win-tie-loss [44] cannot work well in this case because their ranking mechanism depends on the significance test between different methods.
4) An evaluation of ensembles of adjustment methods against single adjustment methods using *SA*, effect size and other ranking methods, to determine whether ensembles are successful in improving performance of single adjustment methods.

In summary, this study is the first work to investigate ensembles of adjustment methods and the first work to create ensembles using Scott-Knott test and Borda count procedure. The remainder of the paper is structured as follows: Section 2 presents an overview of ensemble methods, as well as, the related work on adjustment methods and ensembles in software effort estimation. Section 3 describes the methodology conducted in this research. Section 4 shows the obtained results, which are discussed in Section 5. Section 6 presents threats to validity of our study. Finally, Section 7 presents our conclusions.

## 2 BACKGROUND AND RELATED WORK

### 2.1 ENSEMBLES IN SOFTWARE EFFORT ESTIMATION

Ensembles are learning methods that combine single (aka base) predictive models through a particular aggregation mechanism. The prediction given by the ensemble is a combination of the predictions given by each of its base models, e.g., weighted average [46]. The principal idea of ensembles is that if their models are accurate and diverse, then their performance will be better than the one of its base models. Two models are said to be diverse if they make different errors on the same examples [60]. It is expected that diverse base models will give poor predictions to different examples. So, the poor predictions of a few models can be compensated by the good predictions of others, and the ensemble as a whole can achieve better performance than its base models [59]. On the other hand, if the ensemble is composed of non-diverse base models, its performance will not be better than its base models' individual performances [12][43][50].

The majority of studies in software effort estimation attempt to develop a new estimation method, and then compare the performance of that method against some well-known historical methods under certain conditions [31]. The area of software estimation appears now saturated with many predictive methods. Therefore, rather than developing new methods, there is a trend to replicate previous studies and investigate how we can benefit from their strengths. In practice, measuring accuracies of a particular method against some historical methods under certain settings cannot remain valid when changes on experimental conditions are made [32]. Thus, the method that is being considered superior over dataset $X$ may not remain superior over other datasets or under different parameters [51]. These facts are also true for EBA adjustment methods since most of them use learning methods that need parameters configuration for each training dataset. So, rather than proposing a new adjustment method we aim to benefit from the existing ones by using *Ensembles*. Ensembles have been increasingly used in software engineering to solve regression and classification problems. In the software effort estimation area, Jorgensen recommends that when generating better estimates in expert judgment, it is necessary to use multiple decisions rather than a single one [14].

Kocaguneli et al. [44] distinguish between two main categories of prediction methods: *learner* method and *solo* method. Learner is a single method without supplement of pre or post-processing stages. The *Solo* method is a method supplied with a pre-processing stage such as normalization and/or feature selection. Accordingly, the term *mutli-method* is used to indicate a collection of two or more solo methods [44]. Different solo methods can be used to construct ensemble methods because they present different biases and assumptions. The importance of ranking stability and ensemble methods was studied over 90 solo methods and 20 datasets. The results obtained concluded that the ensemble methods were consistently superior, trustworthy and had a smaller error rate. However, their ensemble method is not guaranteed to work well in other contexts, because it concentrates on selecting the most accurate and stable solo methods, and there is no guarantee that these methods will behave diversely. Using different solo-methods does not guarantee that the corresponding base methods will behave diversely enough, i.e., it does not ensure that they are adequate for composing ensembles. In particular, it is known that there is a trade-off between diversity and accuracy of base models [58]. So, if solo-methods are chosen based on their accuracy only, the ensemble may lack diversity. Therefore, additional studies are necessary when combining other types of solo methods, in order to check whether they would lead to well performing ensembles.

Minku et al. [56] studied the performance of existing automated ensembles and locality approaches on improving accuracy of software effort estimation. The findings obtained show that bagging ensembles of regression trees frequently performed better and, when they did not perform better, their performance was not far from the best one. So, the risk in opting to use bagging ensembles of regression trees instead of one of the other methods is low. They also concluded that combining the power of automated ensembles and locality can lead to competitive results in effort estimation. Similarly, both Pahariya et al. [35] and Kultur et al. [24] also reported significant improvements over single methods, even though these two studies did not provide sufficient information about how the methods' parameters were chosen. Azhar et al. [11] reached to the same conclusion but using web cost estimation models.

On the other hand, other studies report contradictory results and come to different conclusions, such that ensemble methods failed to provide statistically significant improvements on the prediction accuracy. An example of this scenario can be found in the study of Khoshgoftaar et al. [20], [52] in the domain of software quality. They investigated the performance of various ensemble methods combined from 17 learners, over 7 datasets. Findings on single learners outperformed those of ensembles, such that ensembles failed to statistically increase predictive accuracy. A similar conclusion was obtained by both studies conducted by Kocaguneli et al. [21] and Vinaykumar et al. [42].

Kocaguneli et al. [21] failed to improve the predictive performance of ensemble methods under different scenarios. They used various multi methods combined from 14 different effort estimation methods, applied to three datasets. Their study was a replication to Khoshgoftaar et al. [52] study, but in the area of software effort estimation. Vinaykumar et al. [42] investigated two kinds of ensembles combined from various learners, but the obtained results were not generally successful. Braga et al. [57] reported that there was no improvement when using Ensemble methods of Regression Trees and Multiple Linear Perceptron's over single learners. However, these findings were not confirmed by statistical significance tests.

Other studies attempted to develop ensembles from the same learner by either applying pre-processing or post processing, or by changing learner parameters. Examples of this approach are the studies conducted by Twala et al. [39], who used ensemble of imputation methods to handle missing data, and by Wu et al. [54], who used linear combination of CBR methods with different similarity measures. In the same direction, Pahariya et al. [35] investigated multiple variants of genetic algorithms [35], whereas Kultur et al. used collections of neural networks [24].

The main conclusion drawn from these controversial results is that, among other factors, not only the procedure used to evaluate the methods being compared, but also the type of ensemble method being used may influence the conclusions regarding the usefulness of ensembles in software effort estimation. Therefore, it is important to carefully design the ensembles given the problem in hands and to use a principled methodology for evaluation. In this work, we propose a method to consistently and automatically evaluate and select base models to construct ensembles of adjustment methods. Our methodology is well grounded by statistical tests and performance measures able to determine whether the methods being evaluated produce meaningful predictions based on their absolute performance in comparison to random guess [51][55]. Our ensemble approach is carefully designed to only include base methods producing meaningful predictions.

## 2.2 ADJUSTMENT METHODS

Analogy-Based Effort Estimation (EBA) was first introduced informally by Boehm in the beginning of 1980s as a human intensive approach based on opinions of experts [6], in which the expert can use his/her best knowledge to select analogies. This is rather impractical in some circumstances, because there is a danger that an estimator will use an analogy blindly without justifying his/her selection. Later, Shepperd & Schofield [36] established a robust framework for this approach using data-intensive analogy estimation as shown in Figure 1. The basic framework includes four steps: (1) retrieve similar projects, (2) reuse nearest analogies, (3) revise and adjust their solutions, and (4) retain the estimated project in the repository for future use. Shepperd & Schofield [36] described the data in the historical case base as two parts: project description and project solution. Project description is a set of features that define the problem, while project solution defines the effort needed to accomplish the software project [16].

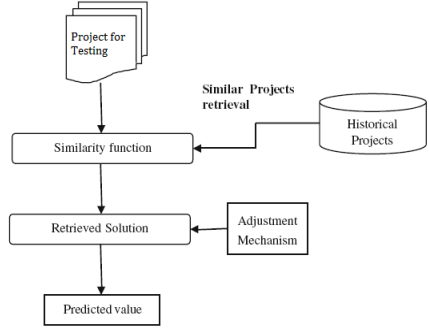

Fig. 1. Process of adjusted EBA method [36]. When the nearest projects are retrieved by similarity function, the process of adjustment begins to calibrate and tune effort of the new project.

Project effort adjustment is the third important stage in the framework above. It can be described as the process of minimizing the difference between new project and retrieved projects, taking into account the characteristics of such projects. The general form of adjustment method can be seen as a simple function as shown in Eq. 1.

$$e_t = adj(x_1, x_2, ..., x_k) \quad (1)$$

where $e_t$ is the final predicted effort of the new project, $adj$ is the adjustment function, and $x_1, x_2 \ldots x_k$ are the nearest projects.

Developing adjustment methods for EBA was the core of several research studies [3][8][19][27][28][41]. Azzeh [4] classified the existing adjustment methods into two main categories according to the procedure they follow: linear and nonlinear. Most of the methods in the literature follow a linear approach except Neural Networks and Model Tree methods, which follow nonlinear adjustment mechanisms. Based on a literature review in the ABE domain, we have found at least 8 different adjustment methods:

- Null Adjustment using conventional EBA [36].
- Linear Size Extrapolation (LSE) [41].
- Multiple Linear Feature Extrapolation (MLFE) [19]
- Regression Towards the Mean (RTM) [15]
- Similarity Based Adjustment (AQUA) [27]
- Model Tree [3].
- Genetic Algorithms based adjustment (GA) [8]
- Neural Networks (NN) [28]

Table 1 summarizes the adjustment methods employed in this study. For each method, a brief description of the method, adjustment model, the adjustment category (none, linear or nonlinear), whether or not categorical features are used for the adjustment, and the features used for the adjustment are shown. It is interesting to note that most methods follow a linear mechanism and some of them do not include categorical features in the adjustment process. Remarkably, the size feature is the only common used feature among all adjustment methods, which is an indication of the importance of the size feature in any estimation method.

LSE was first used by Walkerden and Jeffery [41] in an attempt to tune nearest analogies based on size extrapolation. Although this method was validated over very limited datasets, it showed good performance in comparison with baseline EBA (EBA0). The success of this model depends mainly on the strong correlation between size and effort and the availability of an appropriate size feature. As features other than size may have potential impact on the final estimation, it is believed that other methods considering additional features can be more accurate. This was the core of the studies carried out by Kirsopp et al. [19] and Mendes et al. [30]. They indicated that the arbitrary number of size related features, should be included in the adaption function, this is known as MLFE adjustment strategy. Hence, it seemed to us that a careful use of categorical features should also be considered.

Jorgensen et al. [15] turned the attention to the benefits that can be achieved from some statistical methods such as Regression Towards the Mean (RTM). They claimed that it would seem wise to take advantage of both historical size and effort together in one model. The fundamental concept of RTM assumes that the projects with quite similar structure must be expected to have similar productivity values. Therefore, those projects should be adapted to bring them closer to the average value for the projects in the dataset. Jorgensen et al. [15] reported that the productivity distribution of estimated projects is expected to be narrower than that of actual projects, which leads that the estimated efforts regress towards the mean effort in a particular dataset. To take advantage of this adjustment strategy, it is recommended (but not mandatory) to partition the dataset into more homogeneous and controllable subsets so that the procedure regresses to a local productivity mean [38]. On the other hand, another study showed that the similarity degrees could be used to reflect the amount of changes that will update and adapt the nearest analogies, e.g. AQUA [27]. The certain limitation of this approach is that it is highly affected by similarity measure that may change.

TABLE 1
Summary of adjustment methods

| Method | Adjustment Equation | Adjustment mechanism | Categorical features included? | Adjustment feature |
|---|---|---|---|---|
| EBA | $e_t = \frac{1}{k}\sum_{i=1}^{k} e_i$ (2) | Null | NO | No features, Only effort |
| LSE | $e_t = \frac{1}{k}\sum_{i=1}^{k} \frac{e_i}{size_i} \times size_t$ (3) | Linear | NO | Size (FP or KLOC) |
| MLFE | $e_t = \frac{1}{k}\sum_{i=1}^{k}\left(\frac{1}{m}\sum_{j=1}^{m}\left(\frac{f_{tj}}{f_{ij}} \times e_i\right)\right)$ (4) | Linear | NO | All size related features |
| RTM | $e_t = \frac{1}{k}\sum_{i=1}^{k} size_i \times [pr_i + (h - pr_i) \times (1-c)]$ (5) | Linear | NO | Productivity (size & effort) |
| AQUA | $e_t = \frac{\sum_{i=1}^{k} sim_i \times e_i}{\sum_{i=1}^{k} sim_i}$ (6) | Linear | YES | All relevant features |
| MT | $e_t = \frac{1}{k}\sum_{i=1}^{k}(e_i + d(x_t, x_i))$ (7) | Nonlinear | YES | All relevant features |
| GA | $e_t = \frac{1}{k}\sum_{i=1}^{k}\left(e_i + \sum_{j=1}^{m}\alpha_j \times (f_{tj} - f_{ij})\right)$ (8) | Linear | YES | All relevant features |
| NN | $e_t = \frac{1}{k}\sum_{i=1}^{k}(e_i + NN(R_t, R_k))$ (9) | Nonlinear | YES | All relevant features |

Where
- $e_i$ is the initial effort of $i^{th}$ nearest project.
- $size_i$ and $size_t$ are the project size of $i^{th}$ nearest project and new project.
- $f_{tj}$ and $f_{ij}$ are the $j^{th}$ feature values of the new project and its $i^{th}$ nearest project respectively.
- $m$ is the number of used features.
- $pr_i$ is the productivity of $i^{th}$ nearest analogy which is measured as ($e_i/size_i$),
- $h$ is the average productivity of the similar projects
- $c$ is the historical correlation between the non-adjusted analogy based productivity and the actual productivity as a measure of the expected estimation accuracy.
- $sim_i$ is the similarity value between new project and $i^{th}$ source project.
- $d(x_t, x_i)$ is the Model Tree learning for the effort difference value between new project ($x_t$) and its $i^{th}$ nearest project ($x_i$).
- $NN(R_t, R_k)$ is the Neural Networks training method, $R_t$ is the residuals between new project and the $i^{th}$ project and $R_k$ is the set of residuals between $k$ nearest analogies.

    Genetic algorithm (GA) was also used to tune the difference between a new project and its analogies with respect to all project features [8]. The basic principle of using GA is to optimize the weight coefficient of each feature distance using one objective function. The main difference of this strategy in comparison to others is that it needs too many parameter configurations and user interactions such as chromosome encoding, mutation and crossover. This makes replication a somewhat difficult task. On the other hand, Li et al. [28] used NN to learn the difference in effort values between a new project and its analogies through learning effort differences between historical projects in the training dataset and then the difference produced is added to the nearest analogy.

    Azzeh [3] used Model Tree to adjust and tune selected projects. This adjustment method consists of two stages: Learning and prediction. During the learning phase, the differences between each historical project and its nearest project in the training dataset are computed across all features including efforts. These differences are then used to construct a Model Tree where differences in effort values are considered the output, and differences with respect to features are considered inputs. During the prediction phase, the nearest project to the new project is identified, then differences between them across all features are entered to the constructed Model Tree in order to compute the possible difference in the effort. This amount of difference is

then added to selected effort to produce hopefully better estimate.

The literature on adjustment methods shows that there is no consensus regarding the use of features and their types in addition to the best number of nearest analogies. Each method uses a different set of features, for example: LSE uses only the size feature, MLFE uses the set of size features, RTM uses size and effort features, GA and NN use all features. The accuracy of these adjustment methods is subject to dataset characteristics and distribution of data. Moreover, the results from those studies showed that some adjustment methods cannot outperform conventional EBA over some datasets. So, we believe that rather than creating new adjustment models, it would interesting to investigate ensemble learning to benefit from the advantages of different existing adjustment methods.

## 3 METHODOLOGY AND EXPERIMENT SETUP

### 3.1 Forty Variants of Adjustment Methods

The methods investigated in this study are collection of linear and nonlinear adjustment methods. These methods were selected because their use has previously been examined in the area of effort estimation.

Constructing ensembles favors using different methods that fail under different circumstances [9][23][40]. Specifically, ensemble methods perform better when some members of the ensemble correct the errors made by other members. Each adjustment method used in this study has its own bias and assumptions as they use different statistical or learning methods. For example, making adjustment using neural networks is totally different than using Genetic algorithms.

In addition, each adjustment method requires determination of the number of nearest analogies that will be involved in the procedure of adjustment [53]. Changing $k$ number of nearest analogies makes each variant of the same method behave differently. However, the majority of studies in the field of EBA favor using a pre-determined number of analogies [4]. Kirsopp et al. [19] proposes making predictions based on the 2 nearest projects, as this was found to be the optimum value for the datasets used in their study [4]. In a further study, Kirsopp et al. [19] have improved their accuracy values with case and feature selection algorithms [30]. Similarly, Lipowezky et al. [29], and Walkerden and Jeffery [41] only used the nearest analogy, while Mendes el al. [30] used 1, 2 and 3 nearest analogies. On the other hand, Li et al. [27] made rigorous investigations on public software datasets and noticed different impacts when using various numbers of nearest projects. They also recommended that it is sufficient to use 1 to 5 nearest analogies in order to construct accurate EBA models.

Based on the above findings we can conclude that most EBA methods in literature employed fixed number of analogies ranging from 1 to 5 (i.e. $k \in \{1, 2, 3, 4, 5\}$). Therefore, we adopt this approach and develop five variants for each adjustment method. For example, for RTM method we used RTM1, RTM2…RMT5 methods to represent RTM with one analogy, two analogies and so forth.

### 3.2 Performance measures and statistical test

Performance measures, also known as error measures, are very important indicators for the predictive accuracy of any estimation model. The common accuracy measures that are used to evaluate software effort estimation methods are: 1) Magnitude of Relative Error (*MRE*), which assesses the absolute percentage of error to the actual effort as shown in Eq. 11, 2) Mean of MRE (*MMRE*), as shown in Eq. 12, and 3) Performance indicator (*Pred*(*0.25*)), which is used as a complementary criterion to count the percentage of *MREs* that fall within less than 0.25 of the actual values as shown in Eq. 13.

Despite of the widespread use of these performance measures, there was a substantive discussion about their efficacy for evaluating effort estimation approaches. *MRE* has been criticized for being biased and unbalanced in many validation circumstances, because it yields asymmetry distribution [10], [34], [55]. *MMRE* and *Pred* are based on *MRE*. So, they are also biased measures [55]. To avoid this pitfall, we used the Mean Absolute Error (*MAE*), which is not biased and it does not present asymmetric distribution as *MMRE*. The *MAE* is simply calculated by taking the average of Absolute Error (*AE*) as shown in Eqs. 10 and 14. We also use Standardized Accuracy (*SA*) measure as shown in Eq. 15 and effect size as shown in Eq. 16. The *SA* measure is used mainly to test whether the prediction model in hand really outperforms a baseline of random guessing and generates meaningful predictions. If not so, we cannot even claim that this prediction model is meaningful. *SA* can be interpreted as the ratio of how much better a given model is than random guessing, giving a very good idea of how well the approach does. The effect size ($\Delta$) is used to check whether the predictions of the model in hand are generated by chance, and to justify if there is large effect improvement over random guessing since the statistical significance test alone is not so informative if both predictions models are significantly different. The value of $\Delta$ can be interpreted in terms of the categories of small (0.2), medium (0.5) and large (0.8) where value larger than or equal 0.5 is considered better [55].

Shepperd and MacDonell [55] also recommend using the 5% quantile of the random guessing to estimate the likelihood of non-random estimation. The interpretation of the 5% quantile for random guessing is similar to the use of α for conventional statistical inference that is any accuracy values that is better than this threshold has a less than one in 20% chance of being a random occurrence.

$$AE_i = |e_i - \hat{e}_i| \tag{10}$$

$$MRE_i = \frac{AE_i}{e_i} \quad (11)$$

$$MMRE = \frac{1}{n}\sum_{i=1}^{n} MRE_i \quad (12)$$

$$pred(0.25) = \frac{100}{N} \times \sum_{i=1}^{N} \begin{cases} 1 & if\ MRE_i \leq 0.25 \\ 0 & otherwise \end{cases} \quad (13)$$

$$MAE = \frac{\sum_{i=1}^{n} AE_i}{n} \quad (14)$$

$$SA = 1 - \frac{MAE}{\overline{MAE}_{p_o}} \quad (15)$$

$$\Delta = \frac{MAE - \overline{MAE}_{p_o}}{SP_o} \quad (16)$$

Where:
- $e_i$ and $\hat{e}_i$ are the actual and estimated effort of a particular observation.
- $MAE$ is the mean absolute error of the prediction model.
- $\overline{MAE}_{p_o}$ is the mean value of a large number runs of random guessing. This is defined as, predict a $\hat{e}_i$ for the target case $t$ by randomly sampling (with equal probability) over all the remaining $n - 1$ cases and take $e_t = e_r$ where $r$ is drawn randomly from $1...n \wedge r \neq t$. This randomization procedure is robust since it makes no assumptions and requires no knowledge concerning population.
- $SP_o$ is the sample standard deviation of the random guessing strategy.

In addition to the above mentioned accuracy measures, we used other three accuracy measures mentioned in the literature [44][56] that are considerably less vulnerable to bias or asymmetry distribution as in case of *MMRE*. These measures are Logarithmic Standard Deviation (*LSD*), Mean Balanced Relative error (*MBRE*) and the Mean Inverted Balanced Relative Error (*MIBRE*) as shown in Eqs. 17, 18 and 19 respectively. These measures are primarily used in ranking single and ensemble models as will be explained in Section 3.4.

$$LSD = \sqrt{\frac{\sum_{i=1}^{n}\left(\lambda_i + \frac{s^2}{2}\right)^2}{n-1}} \quad (17)$$

$$MBRE = \frac{1}{N}\sum_{i=1}^{N} \frac{AR_i}{\min(e_i, \hat{e}_i)} \quad (18)$$

$$MIBRE = \frac{1}{N}\sum_{i=1}^{N} \frac{AR_i}{\max(e_i, \hat{e}_i)} \quad (19)$$

where $s^2$ is an estimator of the variance of the residual $\lambda_i$, and $\lambda_i = \ln(e_i) - \ln(\hat{e}_i)$

We also use the Scott-Knott cluster analysis to statistically compare all methods over a specific dataset and then cluster them into homogenous subgroups, where each subgroup contains methods that are significantly indifferent. The Scott-Knott is a multiple comparison statistical procedure based on the notion of clustering where the criterion of clustering is the significance test between methods' absolute error. Since the Scott-Knott presumes provisionally that the error should be normally distributed, we should make sure that all methods' errors are transformed to be normally distributed. For that, we use one of the common transformation methods, namely the Box-Cox method. The Scott-Knott procedure follows and uses one-way analysis of variance (one way ANOVA), which tests the null hypothesis that the methods under comparison are statistically indifferent against the alternative hypothesis that says that the methods can be partitioned into subgroups. The reason behind using Scott-Knott method is its ability to separate the methods into non-overlapping groups. Further details about this test can be found in [51].

**3.3 Method Ranking**
As explained in previous section, Scott-Knott allows us to identify the best methods that have smallest *MAE*. However, when we come to construct ensembles as will be explained in Section 3.4, we should first rank those best methods using not only

*MAE*, but taking other performance measures in consideration. So we used Ranked Voting (RV) method to aggregate ranks across multiple experimental conditions (i.e. performance measures). This method can work well with our assumption since all best methods identified by Scott-Knott are statistically similar. Methods like win-tie-loss [44] cannot work well because their ranking mechanism depends on the significance test between different methods.

This method has never been used before to rank effort estimation models, which is considered part of novelty in this research. RV is a measure of individual interests and preferences as an aggregate towards collective decision [5][25]. The preference order of any voter can be represented by a sequence of candidates in which the first appeared candidate is the top ranked one. For example, the following sequence: (*a*≻*b*≻*c*≻*d*≻*e*) represents the order of 5 candidates where *a* is ranked first then *b*, then *c* and so forth. We should note that the experimental conditions are the voters and adjustment methods acting as possible candidates; so, each voter sorts adjustment methods according to their performances in that experimental condition.

Among many RV methods, we recommend to use Borda counting because it can always be expressed as complete weak preference orders, i.e. they can contain indifferences, but not cycles or intransitivity [26]. In order to calculate the collective decision using Borda, we should first construct a majority margins matrix (MM) as shown in the illustrative example in Table 2. Suppose we have 5 candidates (*a, b, c, d, g*) and 4 experimental conditions (*e1, e2, e3, e4*). Every experimental condition ranks candidates in a definitive order as follows: (*e1: b*≻*a*≻*d*≻*c*≻*e, e2: a*≻*d*≻*b*≻*c*≻*e, e3: b*≻*d*≻*a*≻*e*≻*c, e4: a*≻*d*≻*e*≻*b*≻*c*). Each entry in MM represents how many times a candidate *x* proceeds candidate *y* across all performance measures. This can be accomplished by subtracting the times that *x* beats *y* (|*x*≻*y*|) from the times that *y* beats *x* (|*y*≻*x*|). For example, the first row and third column tell us that MM*a,c* = |*a*≻*c*| -|*c*≻*a*| = 4-0=4, which indicates that the candidate *a* beats the candidate *c* by a margin of four. After that, the summation of votes for every candidate over each condition is calculated. The candidates are then ranked in a descending order based on the final summation where the candidate with the largest score is the top winner.

TABLE 2
Majority margin matrix where rows and columns headers represent possible candidates. Each entry represents precedence degree between two candidates across all voters. The last column represents the overall score that candidates are sorted upon.

|   | a  | b  | c | d  | g | Score |
|---|----|----|---|----|---|-------|
| a |    | 0  | 4 | 2  | 4 | **10** |
| b | 0  |    | 4 | 0  | 2 | **6** |
| c | -4 | -4 |   | -4 | 0 | **-12** |
| d | -2 | 0  | 4 |    | 4 | **6** |
| g | -4 | -2 | 0 | -4 |   | **-10** |

The resulted aggregated scores for every candidate are represented in bold in the last column of Table 2. Therefore, for the above profile we get the following ranking: *a*≻(*b*~*d*)≻*g*≻*c* where candidates on the left hand side are ranked higher and ~ symbol means indifference between two candidates (i.e. they have same rank). The final ranking suggests that the candidate *a* is the top winner.

### 3.4 Building Ensembles
In this section we describe the procedure used to construct ensembles from variant of adjustment methods. In the first step we evaluate all EBA variants to check whether they are predicting well using *SA* and effect size accuracy measures. If any method fails to show improvement with respect to random guessing or show small effect size improvements, it is ignored from further experiments. In other words, any method to pass this test should obtain larger *MAE* than the *MAE* of random guessing and effect size greater than 0.5 (medium effect size). Methods that pass this test are considered to be actual prediction methods. This step is conducted for all methods over each dataset individually.

All actual prediction methods are evaluated in terms of *AE* and *MAE* and clustered based on the significance test of Scott-Knott method. The best subgroup's methods are then selected as potential ensemble members to the corresponding dataset, as they have potential to produce more accurate ensembles if they are combined together. Before constructing ensembles we consult Borda count to rank the selected best methods from Scott-Knott test based on various accuracy measures: *MAE, LSD, MBRE* and *MIBRE*. So, in this case, we compute error measures for every selected method over each dataset, and then sort them according to their accuracies. The accumulative ranking is obtained by applying Borda count.

Based on the obtained rankings, we start constructing ensembles from top variants according to the position they occupy, in this manner: *Top2, Top3 TopZ…TopM*, where *TopZ* represents the ensemble method that is being constructed from the first *Z* ranked methods. The predictions made by an ensemble are the average predictions of its methods. Finally, like single methods, the ensemble methods are evaluated over all datasets using *SA* and effect size, and compared to other single methods using Scott-Knott and Borda count. All single methods and ensembles are ranked to identify superior methods.

### 3.5 Experiment setup
Choosing and setting experiment conditions is important for the validity of any method [4][33]. In this section, we describe the experimental conditions employed in this study. All single methods and ensembles should undergo the same experimental condition in order to have consistent results regarding their predictive accuracy.

Each method is evaluated over all datasets using leave-one-out cross validation to identify test and train data. In each run,

one project is selected for testing while the remaining projects are used for training. The training data are used to construct a method while test data are used to validate the method. This procedure is performed until all projects within a dataset are used as test projects.

Concerning the distance measure that will be used to retrieve nearest projects, we used the most common measure, which is the un-weighted Euclidian distance as shown in Eq. 20. Note that the distance between categorical values is computed as binary, i.e. 1 if they are different and 0 otherwise.

$$d = \sqrt{\sum_{i=1}^{m}(x_i - y_i)^2} \tag{20}$$

where $d$ is the distance between projects $x$ and $y$ across $m$ features.

Since all continuous features in any dataset have different ranges, they influence the similarity degree relatively. Therefore, all continuous features were normalized using min-max approach to be in the range between 0 and 1 [22] when selecting the nearest projects. Note that the normalization was used only for selecting nearest projects, and not within the adjustment method itself. This is because some methods such as LSE and RTM require to use certain features such as size as a denominator (see table 1), and features with value zero cannot be used as denominator.

## 3.6 Research Methodology
Figure 2 shows the flow chart of the research methodology of this article, which is explained by the following steps:
1- We used the accuracy measures and validation framework proposed by Shepperd and MacDonell [55] as explained in Section 3.2 to test whether the adjusted EBA variant is actually predicting or it generates its predictions by chance. The prediction methods that show positive improvements in *SA* in comparison to random guessing and have effect size greater than the suggested medium value (i.e. 0.5) are selected as potential contributors in the ensembles. The effect size measure is very informative to give us the level of improvements especially in the case of significant difference between two prediction methods. In this case, the effect size serves as decision maker to significantly stabilize the ranking of the existing adjustment methods based on their *MAE* accuracy measure.
2- After determining the adjustment methods that lead to meaningful predictions, we use Scott-Knott test method to significantly cluster and select the best adjustment methods that have smallest *MAE*.
3- The best adjustment methods are ranked using Borda count across 4 error measures (*MAE*, *LSD*, *MBRE*, *MIBRE*), being applied for each dataset individually. This allows us to construct ensembles in an automated way.
4- Build and evaluate ensembles from the top best single methods as mentioned in Section 3.4.

## 3.7 Datasets
In order to assess the performance of any method, it is necessary to validate such method over some historical datasets that exhibit different characteristics. Most of the methods in literature were tested on a single or a very limited number of datasets, thereby reducing the credibility of the proposed method [13]. To avoid this pitfall, we included 8 public software effort datasets that come from different industrial sectors. Specifically, these datasets come from PROMISE repository [7] which is an on-line publically available data repository and it consists of datasets donated by various researchers around the world. The datasets come from this source are: Desharnais, Kemerer, Albrecht, Cocomo, Maxwell, China, Telecom and Nasa datasets. The employed datasets typically contain a unique set of features that can be categorized according to four classes [45]: size features, development features, environment features and project data. Table 3 shows the descriptive statistics of such datasets. From these statistics we can conclude that datasets in the area of software effort estimation share relatively common characteristics [17]. They often have a limited number of observations that are affected by multicollinearity and outliers. Notably, all datasets have positive skewness efforts that range from 1.78 to 4.36 which indicate that the effort of each dataset is not normally distributed and presents a challenge for developing accurate estimation methods.

TABLE 3
Statistical properties of the employed dataset

| Dataset | Feature | Number of Projects | Effort Data | | | | | |
|---|---|---|---|---|---|---|---|---|
| | | | unit | min | max | mean | median | Skew |
| Albrecht | 7 | 24 | months | 1 | 105 | 22 | 12 | 2.2 |
| Kemerer | 7 | 15 | months | 23.2 | 1107.3 | 219.2 | 130.3 | 2.76 |
| Nasa | 3 | 18 | months | 5 | 138.3 | 49.47 | 26.5 | 0.57 |
| Desharnais | 12 | 77 | hours | 546 | 23940 | 5046 | 3647 | 2.0 |
| Cocomo | 17 | 63 | months | 6 | 11400 | 683 | 98 | 4.4 |
| China | 18 | 499 | hours | 26 | 54620 | 3921 | 1829 | 3.92 |
| Maxwell | 27 | 62 | hours | 583 | 63694 | 8223.2 | 5189.5 | 3.26 |
| Telecom | 3 | 18 | months | 23.54 | 1115.5 | 284.33 | 222.53 | 1.78 |

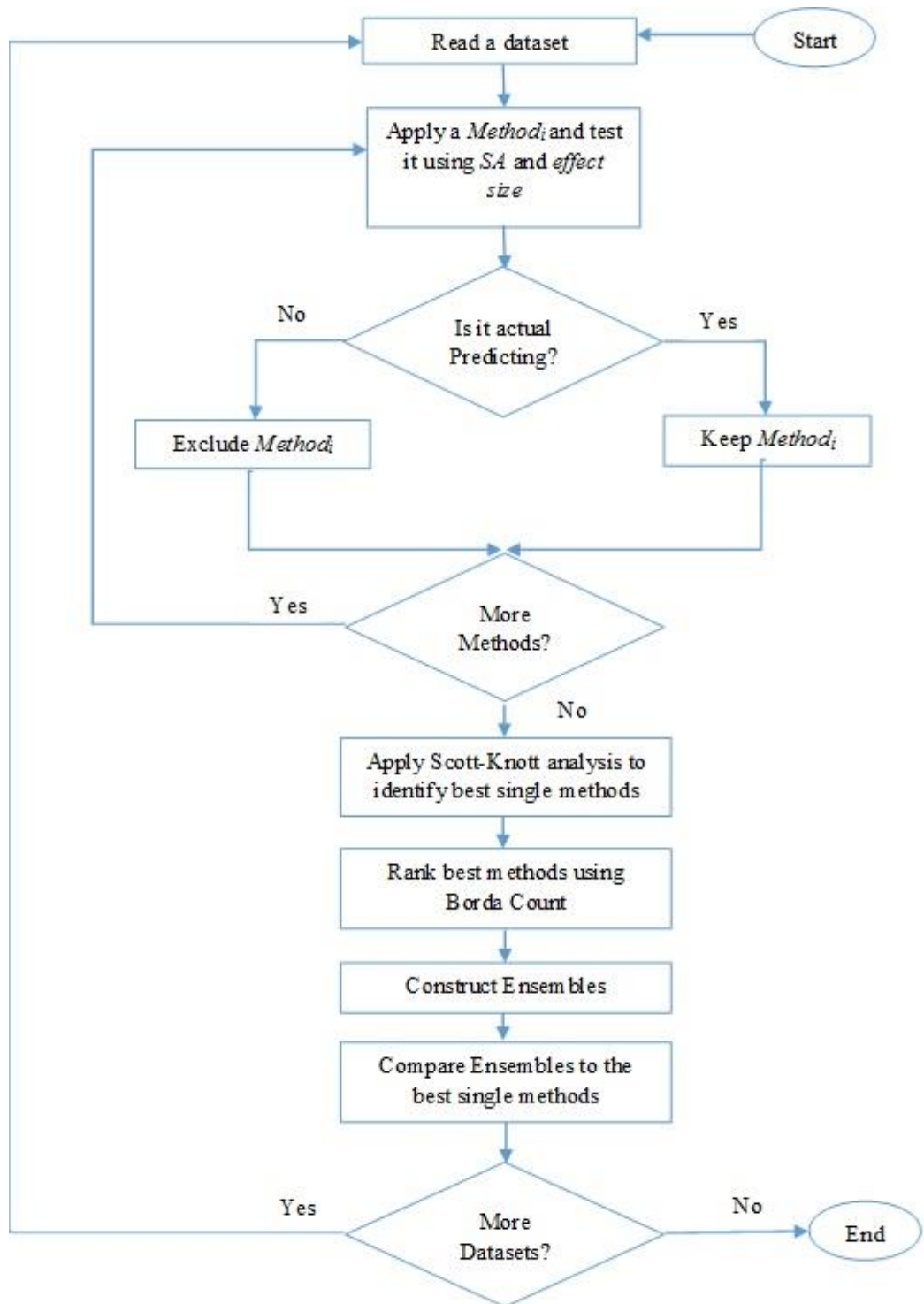

Fig. 2. Flow chart describing the research methodology explained in section 3.6

## 4 RESULTS

This section presents the results of the experiments conducted on 8 datasets and 40 adjustment methods with the aim of providing a better understanding of the relationship between datasets, adjustment methods and number of nearest analogies. In the first section, we evaluate the validity of these adjustment methods and their ability to provide actual predictions. Then, we evaluate the constructed ensemble methods against single methods.

### 4.1 Evaluation of Single Adjustment Methods

As a first step of the evaluation, all 40 single adjustment methods have been evaluated over all datasets, using *SA* and effect size to check whether they produce meaningful predictions (i.e., considerably better than random guessing), as explained in Section 3. Based on the obtained findings we select only the methods that present large *SA* and their effect size is greater than 0.5 (i.e. not generated by chance). At this stage we should also ensure that the obtained *MAE* for each method over every dataset falls comfortably beyond the 5% quantile of random guessing. This will help us to make sure that the results are highly unlikely arisen by chance. Table 4 gives the summary of *SA* and effect size for all EBA variants over all datasets where the baseline method used here is the random guessing. In Table 4, we can see that each row reports the *SA* and $\Delta$ accuracy statistics of the adjusted EBA variants. We also provide the *SA* of 5% quantile of random guessing as shown in the second row as *SA5%*.

The accuracy *SA* indicates the relative improvements or otherwise from random guessing and thus it is immediately clear that all methods perform comparatively well and generate considerably meaningful predictions for all datasets with an exception of China dataset where none of the MLFE variants generated better predictions than random guessing. This happens when the *MAE* of random guessing is much smaller than the *MAE* of the MLFE variants. So, they failed to act as prediction methods. If we look at effect size test we can comfortably confirm that the all EBA variants (except MLFE over china) are predicting considerably well, since they yield considerably better accuracy level than random guessing and lie beyond the 5% quantile. The effect size test shows considerably large effect size overall datasets, which confirms large effect improvement over guessing (i.e. $\Delta > 0.5$). All methods that could not pass this test were excluded from further experimentations due to their poor predictions.

## 4.2 Ranking of Single Adjustment Methods

All methods that have passed the abovementioned test were then selected for Scott-Knott cluster analysis in order to identify homogeneous groups of methods and select the group with smallest *MAE* (i.e., the best group). We tested the distribution of *AEs* from all prediction methods using Kolmogorov–Smirnov test, and found that they were not normally distributed. Since the Scott-Knott algorithm is based on the assumption that the distribution of errors is approximately normal, the absolute errors for all methods were transformed, as explained in section 3.2, for being used as input to the Scott-Knott test. Thus ranking was based on the means of the transformed absolute errors, not on the original values. Figures 3(a) to 3(h) show graphical plot of Scott-Knott cluster analysis based on Anova significance test and using transformed absolute error. The x-axis represents comparative method names sorted according to their ranks where better places start from right hand side. The y-axis represents the transformed absolute errors and the small circles on each vertical line represent mean of transformed absolute errors. For example, in Figure 3(a) we present the results for Albrecht dataset. The Scott-Knott analysis resulted in four homogeneous clusters where methods in each cluster show statistically similar performance. We can observe that all datasets have a significantly best group. The gray box on the right most of each figure shows the best subgroup of methods that give significantly smallest *MAE*. Specifically, we can notice that each dataset favors different prediction methods, confirming our previous belief that the choice of adjustment method is often dataset dependent [53].

The general findings that can be extracted from this figure suggest that (1) there is no consistency in the number of methods that belong to the best group for each dataset, (2) there is no stable conclusion regarding the choice of best *k* nearest number of analogies for each dataset and (3) non-linear methods such as NN and MT are rarely selected by the Scott-Knott test as the best methods. If we look around, we can find many factors that affect the performance of adjustment methods such as dataset characteristics, number of projects, types of features, etc. Another note is the ranking instability across all datasets. It is clear that there was no stable ranking across all datasets. If we look closer at the first rank in each dataset, it is impossible to identify any common method between them, so we cannot claim that there is an outperformer across all datasets. However, although there is ranking instability, it is crucial to obtain the important information that is hidden in this lack of convergence and to interpret the findings with caution. The main finding is that, as there is insignificant difference among the predictors in the same group, the expert can use the simplest method to avoid any possible complexity that may be caused by some methods' configuration, as in case of GA and NN. By summarizing the results, we can figure out the following interesting findings:

1) For the datasets with few features such as Nasa and Telecom, there is a trend to favor many methods in the same best subgroup.
2) The most top ranked methods with smallest *MAE* belong to LS and RTM variants, with 11 selections each. This a proof for the importance of the size feature in adjusting software projects, because it carries useful information such as project functionality and complexity.
3) NN and MLFE variants were the worse methods across all datasets, given that they have not been selected in any best subgroup except NN1 for Kemerer dataset.
4) Concerning number of nearest analogies, there is no stable conclusion about which best *k* is the most favored one. However we can see that *k*=1 and 5 were the most dominant choices across all datasets. Even then, we cannot judge that these values are the optimum values because it infeasible to include all possible *k* values, and there is previous experience that the choice of best *k* value is a dataset dependent [53]. For example, for datasets with large number of categorical features, the choice of *k*=1 was the dominant across all methods. As result we can see the performance of Adjusted EBA is never improved by increasing *k* value.

TABLE 4
The *SA* and effect size of 40 adjustment methods used in EBA are listed based on comparison with random guessing baseline method.

|  | Albrecht | | Kemerer | | Desharnais | | Cocomo | | Maxwell | | China | | Nasa | | Telecom | |
|---|---|---|---|---|---|---|---|---|---|---|---|---|---|---|---|---|
| $SA_{5\%}$ | 8.4 | | 14.9 | | 17.5 | | 20 | | 24.2 | | 26.2 | | 27.9 | | 28.3 | |
| | *SA* | Δ | *SA* | Δ | *SA* | Δ | *SA* | Δ | *SA* | Δ | *SA* | Δ | *SA* | Δ | *SA* | Δ |
| EBA1 | 63 | 3.10 | 50 | 1.92 | 37 | 3.92 | 49 | 2.67 | 10 | 0.72 | 66 | 12.58 | 65 | 3.80 | 41 | 2.27 |
| EBA2 | 68 | 3.36 | 45 | 1.73 | 46 | 4.79 | 46 | 2.46 | 30 | 2.21 | 67 | 12.78 | 70 | 4.04 | 49 | 2.72 |
| EBA3 | 67 | 3.29 | 44 | 1.70 | 47 | 4.98 | 44 | 2.40 | 34 | 2.54 | 67 | 12.65 | 66 | 3.81 | 51 | 2.85 |
| EBA4 | 62 | 3.06 | 40 | 1.53 | 50 | 5.20 | 46 | 2.52 | 32 | 2.41 | 65 | 12.32 | 64 | 3.70 | 47 | 2.62 |
| EBA5 | 62 | 3.03 | 39 | 1.51 | 53 | 5.58 | 48 | 2.62 | 31 | 2.33 | 63 | 11.90 | 62 | 3.57 | 44 | 2.45 |
| MT1 | 63 | 3.10 | 50 | 1.92 | 37 | 3.92 | 49 | 2.67 | 10 | 0.72 | 66 | 12.58 | 65 | 3.80 | 41 | 2.27 |
| MT2 | 68 | 3.34 | 47 | 1.82 | 45 | 4.69 | 50 | 2.73 | 26 | 1.93 | 68 | 12.95 | 69 | 4.02 | 47 | 2.62 |
| MT3 | 69 | 3.37 | 47 | 1.81 | 47 | 4.89 | 48 | 2.58 | 31 | 2.31 | 68 | 12.90 | 69 | 4.02 | 49 | 2.75 |
| MT4 | 66 | 3.26 | 45 | 1.71 | 48 | 5.07 | 47 | 2.57 | 33 | 2.44 | 67 | 12.76 | 68 | 3.93 | 49 | 2.75 |
| MT5 | 65 | 3.20 | 43 | 1.64 | 51 | 5.31 | 48 | 2.59 | 33 | 2.45 | 66 | 12.55 | 68 | 3.95 | 49 | 2.72 |
| AQUA1 | 63 | 3.10 | 50 | 1.92 | 37 | 3.92 | 49 | 2.67 | 10 | 0.72 | 66 | 12.58 | 65 | 3.80 | 41 | 2.27 |
| AQUA2 | 67 | 3.31 | 43 | 1.64 | 46 | 4.79 | 43 | 2.31 | 28 | 2.10 | 67 | 12.63 | 70 | 4.08 | 49 | 2.72 |
| AQUA3 | 66 | 3.24 | 41 | 1.56 | 47 | 4.92 | 43 | 2.31 | 33 | 2.47 | 66 | 12.47 | 54 | 3.15 | 48 | 2.70 |
| AQUA4 | 61 | 2.99 | 36 | 1.39 | 49 | 5.14 | 46 | 2.48 | 31 | 2.32 | 63 | 12.04 | 50 | 2.91 | 45 | 2.52 |
| AQUA5 | 60 | 2.95 | 37 | 1.41 | 53 | 5.56 | 48 | 2.60 | 30 | 2.23 | 61 | 11.58 | 50 | 2.91 | 35 | 1.95 |
| LS1 | 70 | 3.44 | 51 | 1.95 | 45 | 4.77 | 58 | 3.13 | 42 | 3.09 | 80 | 15.14 | 83 | 4.84 | 46 | 2.58 |
| LS2 | 74 | 3.62 | 53 | 2.03 | 57 | 5.99 | 69 | 3.72 | 51 | 3.80 | 81 | 15.43 | 89 | 5.15 | 46 | 2.57 |
| LS3 | 73 | 3.58 | 56 | 2.16 | 54 | 5.72 | 62 | 3.38 | 52 | 3.85 | 79 | 14.97 | 89 | 5.17 | 48 | 2.67 |
| LS4 | 69 | 3.39 | 55 | 2.11 | 54 | 5.62 | 66 | 3.56 | 53 | 3.95 | 80 | 15.21 | 91 | 5.26 | 50 | 2.80 |
| LS5 | 70 | 3.41 | 55 | 2.11 | 56 | 5.85 | 67 | 3.63 | 55 | 4.09 | 81 | 15.34 | 91 | 5.28 | 48 | 2.68 |
| MLFE1 | 76 | 3.73 | 51 | 1.95 | 43 | 4.55 | 52 | 2.79 | 42 | 3.09 | -242 | -45.88 | 79 | 4.57 | 43 | 2.39 |
| MLFE2 | 67 | 3.31 | 50 | 1.91 | 51 | 5.31 | 48 | 2.59 | 51 | 3.80 | -181 | -34.42 | 81 | 4.70 | 44 | 2.47 |
| MLFE3 | 70 | 3.46 | 48 | 1.84 | 50 | 5.29 | 47 | 2.55 | 52 | 3.85 | -175 | -33.15 | 83 | 4.81 | 46 | 2.56 |
| MLFE4 | 70 | 3.42 | 45 | 1.73 | 52 | 5.47 | 49 | 2.67 | 53 | 3.95 | -237 | -44.95 | 80 | 4.63 | 50 | 2.78 |
| MLFE5 | 70 | 3.43 | 47 | 1.81 | 55 | 5.83 | 51 | 2.77 | 55 | 4.09 | -350 | -66.43 | 80 | 4.62 | 47 | 2.61 |
| RTM1 | 68 | 3.36 | 57 | 2.18 | 52 | 5.50 | 60 | 3.26 | 52 | 3.88 | 80 | 15.14 | 88 | 5.12 | 47 | 2.62 |
| RTM2 | 68 | 3.33 | 57 | 2.19 | 57 | 5.98 | 65 | 3.53 | 54 | 3.99 | 81 | 15.41 | 91 | 5.26 | 48 | 2.70 |
| RTM3 | 67 | 3.31 | 57 | 2.17 | 55 | 5.83 | 61 | 3.30 | 52 | 3.89 | 79 | 14.94 | 90 | 5.23 | 49 | 2.73 |
| RTM4 | 65 | 3.19 | 57 | 2.18 | 54 | 5.71 | 64 | 3.44 | 52 | 3.87 | 80 | 15.17 | 90 | 5.21 | 52 | 2.90 |
| RTM5 | 65 | 3.22 | 57 | 2.18 | 56 | 5.83 | 65 | 3.50 | 53 | 3.91 | 81 | 15.29 | 90 | 5.20 | 49 | 2.75 |
| GA1 | 73 | 3.57 | 50 | 1.92 | 37 | 3.93 | 49 | 2.67 | 10 | 0.72 | 66 | 12.58 | 67 | 3.86 | 41 | 2.27 |
| GA2 | 73 | 3.58 | 45 | 1.74 | 46 | 4.80 | 46 | 2.47 | 30 | 2.21 | 67 | 12.79 | 71 | 4.10 | 48 | 2.71 |
| GA3 | 74 | 3.63 | 44 | 1.70 | 47 | 4.98 | 44 | 2.40 | 34 | 2.54 | 67 | 12.65 | 67 | 3.87 | 51 | 2.84 |
| GA4 | 68 | 3.35 | 40 | 1.54 | 50 | 5.20 | 46 | 2.52 | 32 | 2.41 | 65 | 12.32 | 65 | 3.75 | 47 | 2.62 |
| GA5 | 66 | 3.26 | 40 | 1.52 | 53 | 5.58 | 48 | 2.62 | 31 | 2.33 | 63 | 11.90 | 63 | 3.65 | 44 | 2.46 |
| NN1 | 51 | 2.49 | 41 | 1.57 | 38 | 3.98 | 49 | 2.64 | 10 | 0.75 | 67 | 12.66 | 70 | 4.06 | 37 | 2.09 |
| NN2 | 59 | 2.91 | 40 | 1.55 | 46 | 4.83 | 45 | 2.42 | 29 | 2.17 | 68 | 12.83 | 50 | 2.90 | 38 | 2.12 |
| NN3 | 61 | 2.97 | 47 | 1.80 | 48 | 5.02 | 44 | 2.41 | 34 | 2.52 | 67 | 12.70 | 77 | 4.44 | 47 | 2.63 |
| NN4 | 61 | 2.99 | 33 | 1.28 | 50 | 5.21 | 47 | 2.53 | 33 | 2.43 | 65 | 12.36 | 69 | 4.01 | 45 | 2.49 |
| NN5 | 67 | 3.31 | 35 | 1.34 | 53 | 5.59 | 49 | 2.64 | 31 | 2.32 | 63 | 11.92 | 64 | 3.71 | 43 | 2.41 |

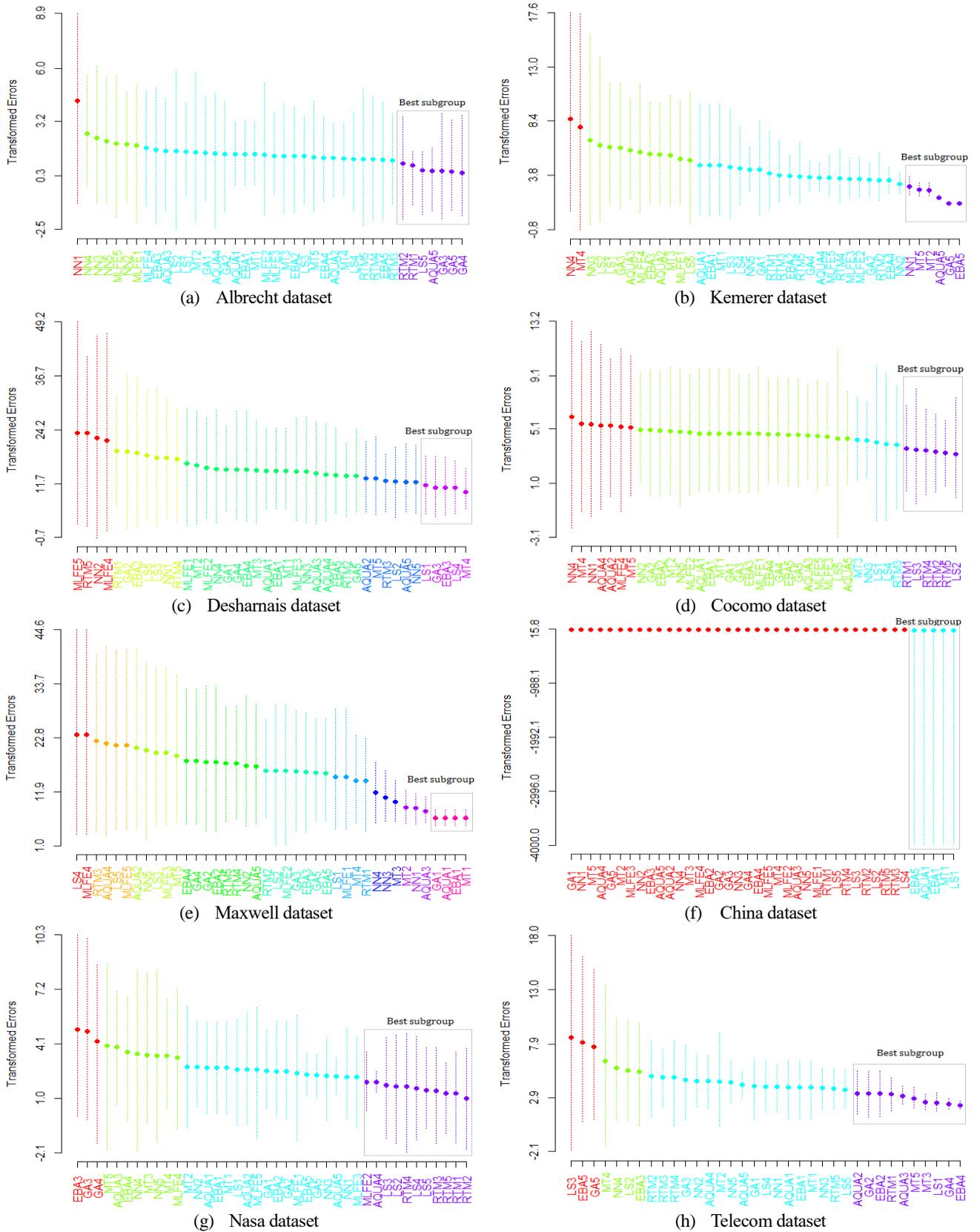

Fig 3. Plot of the Scott-Knott algorithm based on Transformed absolute errors

From now on, we will use only the methods that were in the best group of each dataset, because we believe that these methods are promising since they present smallest *MAE*. That is, they have potential to deliver better accuracy if they work together. However, since all methods in the best group are statistically similar, we cannot rely only on the ranking obtained by Scott-Knott as shown in Figure 3, but we also consult Borda count method to rank all variants across all error measures *MAE*, *LSD*, *MBRE*, *MIBRE*, for each dataset. Table 5 shows adjustment methods variants sorted by the calculated Borda score seen in all error measures over each dataset. The adjustment method with the largest score is ranked #1. At the other end of the scale, the adjustment method with the lowest score is ranked last. Note that all methods are ranked in ascending order (i.e. lowest first) over all error measures. Unlike the Scott-Knott results, it is clear from the final scores that GA, MT and LSE variants tend to rank top after computing the aggregated Borda scores across all error measures, over all datasets. Although the RTM method is one of the accurate adjustment methods for EBA as confirmed in [4], its variants occupied the last positions with the worst performance across all error measures. On contrast, the good performance of LSE variants can be attributed to the fact that this method uses only the size feature as the adjustment factor, given that this feature is considered the strongest correlated feature with effort in all datasets.

In order for a method to be considered superior to others, it should not only be ranked first, but also have a minimum number of changes in their ranks. Therefore, we used average change of ranks (ξ) to measure stability of ranking for each method across different experimental conditions. The average change of ranks is measured by computing differences between all ranks that are obtained for a particular method. Then, we calculate the average of these differences. Smaller average change of ranks represents more stable ranking. From Table 5, we can observe that the top ranked methods have lower rank changes than other methods. This represents a good behavior, showing that the top ranked methods can be considered superior not only in terms of average rank, but also in terms of rank stability. Thus, these superior single methods will be used to form various ensembles based on the assumption made in section 3.

TABLE 5
The best adjustment methods used in EBA are sorted based on scores obtained by Borda counting rule. These methods have been compared over 8 datasets using 4 error measures.

| | Albrecht | | Kemerer | | Desharnais | | Cocomo | | Maxwell | | China | | Nasa | | Telecom | |
|---|---|---|---|---|---|---|---|---|---|---|---|---|---|---|---|---|
| Rank | Method | ξ | Method | ξ | Method | ξ | Method | ξ | Method | ξ | Method | ξ | Method | ξ | Method | ξ |
| 1 | GA5 | 1.5 | MT5 | 1 | MT4 | 1.5 | RTM2 | 1.75 | GA1 | 1 | LS1 | 1.25 | LS3 | 2 | GA2 | 1.75 |
| 2 | GA3 | 2.25 | MT2 | 2.75 | EBA3 | 2.25 | RTM5 | 1.75 | MT1 | 2.25 | EBA5 | 2.25 | LS5 | 2.5 | EBA2 | 2.25 |
| 3 | GA4 | 3 | NN1 | 3.75 | LS1 | 3.25 | LS2 | 3.25 | EBA1 | 2.75 | MT1 | 2.5 | LS2 | 3.5 | AQUA3 | 2.75 |
| 4 | LS5 | 3.25 | EBA5 | 4.25 | GA3 | 3.25 | LS3 | 3.5 | AQUA1 | 4 | EBA1 | 4.5 | RTM1 | 3.5 | RTM1 | 4.5 |
| 5 | AQUA5 | 5.25 | AQUA5 | 5 | LS4 | 4.75 | RTM4 | 4.75 | | | AQUA1 | 4.5 | RTM3 | 5.5 | LS1 | 5 |
| 6 | RTM1 | 5.75 | GA5 | 5.75 | | | RTM1 | 6 | | | | | LS4 | 4.75 | MT5 | 4.75 |
| 7 | | | | | | | | | | | | | RTM2 | 7 | AQUA2 | 6.5 |
| 8 | | | | | | | | | | | | | RTM4 | 8.25 | MT3 | 8.75 |
| 9 | | | | | | | | | | | | | RTM5 | 8.75 | EBA4 | 8.75 |
| 10 | | | | | | | | | | | | | AQUA4 | 9 | GA4 | 9 |

### 4.3 Evaluation of Ensemble Methods

Being able to construct ensembles from all possible combinations of estimation methods is a daunting process (for example for $n$=40 variants of adjustment methods there are $\sum_{r=n}^{2}\left(n!\times((n-r)!r!)^{-1}\right)$ possible ensemble methods). Dealing with this large number of methods is very difficult and time consuming. Therefore we suggest first to identify the methods that most often ranked higher and then construct ensembles from them. The top ranked methods from Table 5 are taken to construct various ensemble methods. The reason behind this choice was because these methods have stable ranking and better performance than others according to the Scott-Analysis presented in Figure 3. It is important to note that we did not take all possible combinations but we suggest building ensembles that are meaningful and realistic. Therefore, we build ensembles from *Top2, Top3…TopM* single methods where *M* is the number of selected methods in each dataset. For each variant of ensemble, the methods are aggregated by the mean of their predictions. Based on this assumption we obtain *M-1* ensembles for each dataset. For example, the *Top2* for albrecht dataset is constructed from both GA5 and GA3 while the ensemble *Top3* is constructed from GA5, GA3 and GA4, and so forth.

TABLE 6
The SA and effect size of ensemble adjustment methods used in EBA are listed based on comparison with random guessing baseline method.

|  | Albrecht | | Kemerer | | Desharnais | | Cocomo | | Maxwell | | China | | Nasa | | Telecom | |
|---|---|---|---|---|---|---|---|---|---|---|---|---|---|---|---|---|
| $SA_{5\%}$ | 8.4 | | 14.9 | | 17.5 | | 20 | | 24.2 | | 26.2 | | 27.9 | | 28.3 | |
|  | SA | Δ | SA | Δ | SA | Δ | SA | Δ | SA | Δ | SA | Δ | SA | Δ | SA | Δ |
| Top2 | 71.20 | 3.86 | 47.29 | 1.86 | 48.32 | 5.26 | 64.81 | 3.54 | 9.60 | 0.76 | 76.58 | 14.40 | 90.29 | 5.03 | 48.62 | 2.73 |
| Top3 | 70.61 | 3.83 | 47.01 | 1.85 | 49.94 | 5.44 | 66.55 | 3.64 | 9.59 | 0.76 | 75.31 | 14.16 | 89.80 | 5.00 | 49.41 | 2.77 |
| Top4 | 70.99 | 3.85 | 46.24 | 1.82 | 49.86 | 5.43 | 65.52 | 3.58 | 9.59 | 0.76 | 73.64 | 13.85 | 89.91 | 5.01 | 49.25 | 2.76 |
| Top5 | 68.95 | 3.74 | 44.46 | 1.75 | 51.28 | 5.58 | 65.16 | 3.56 | | | 72.36 | 13.61 | 90.16 | 5.03 | 49.07 | 2.75 |
| Top6 | 69.27 | 3.76 | 43.72 | 1.72 | | | 64.67 | 3.53 | | | | | 90.30 | 5.03 | 48.50 | 2.72 |
| Top7 | | | | | | | | | | | | | 90.47 | 5.04 | 48.64 | 2.73 |
| Top8 | | | | | | | | | | | | | 90.63 | 5.05 | 48.69 | 2.73 |
| Top9 | | | | | | | | | | | | | 90.80 | 5.06 | 48.74 | 2.74 |
| Top10 | | | | | | | | | | | | | 89.91 | 5.03 | 49.21 | 2.76 |

Similarly, we evaluate each ensemble method in every dataset using *SA* and effect size (i.e. using Leave-one-out cross validation). Like single methods, all ensemble methods are applied to 8 datasets using 4 error measures. The obtained results in Table 6 confirm that all constructed ensemble methods are reasonable predictors, as they fall comfortably beyond 5% quantile of random guessing. Also we can notice that the values of *SA* and effect size for ensemble methods are slightly better than those of single methods, but this need to be backed up by statistical tests. Therefore, the best single methods and ensemble methods are re-sorted over all datasets using Scott-Knott cluster analysis test to statistically identify best performers with smallest *MAE* as shown in Figures 4(a) to 4(h). The main conclusion from this figure is that there is no sufficient evidence to the superiority of ensemble methods over single methods. For example, we cannot find any significance difference between ensemble methods and single methods based on the transformed *AE* for four datasets namely: Albrecht, Maxwell, China and Nasa. This suggests that using single methods could be sufficient and there is no need to construct ensembles. For Kemerer, Desharnais and Cocomo, some ensemble methods can be surpassed by single methods. These findings yield a concern about the accuracy of ensembles in adjusting EBA. Since Top methods are almost always within the group of best methods and as there is no statistically significant difference among methods within a given best group, one cannot conclude that top or single methods within that group were better than each other. To heuristically investigate on this issue further, we consult Borda count to tell us more about this finding.

The obtained rankings after applying Borda counting using *MAE*, *LSD*, *MBRE*, *MIBRE* error measures are shown in Table 7. Surprisingly, all ensemble methods are often ranked higher with highest scores among other single methods. Also, these methods have smallest average change of ranks compared to other methods. It is interesting to note that most ensembles have the smallest average change of ranks seen in any method, and it is rare to see that they occupy last ranks. This result shows that ensembles of top-ranked adjustment methods not only are among the methods with best score, but also obtain generally stable ranking across all experimental conditions. Furthermore, all ensemble methods have relatively stable ranking compared to other single methods. Although the rankings of ensemble methods based on 4 error measures clearly portray an overview of their predictive performance while Scott-Knott test shows the opposite, it is essential to recall the conclusions drawn from previous studies that using various measures yield different rankings. Therefore, according to the results in hands, we can conclude that ensemble of adjustment methods can work well with good performance, but are not always better than single methods.

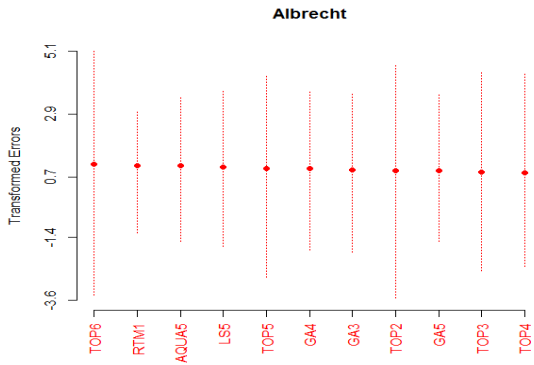
(a) Albrecht dataset

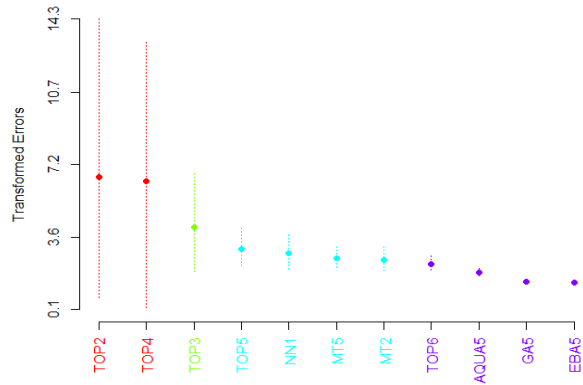
(b) Kemerer dataset

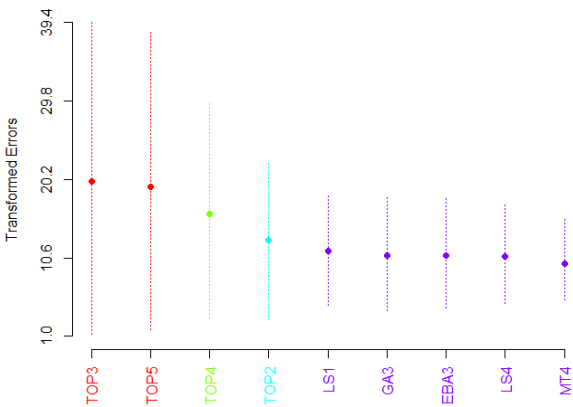
(c) Desharnais dataset

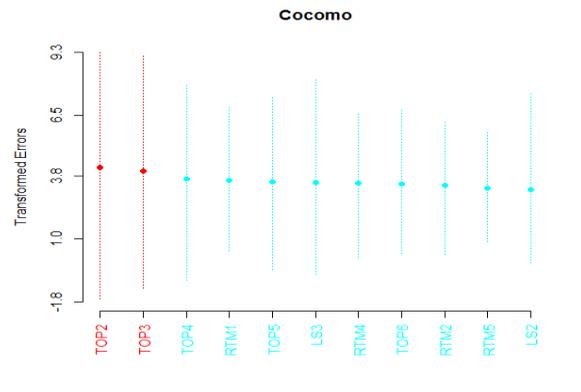
(d) Cocomo dataset

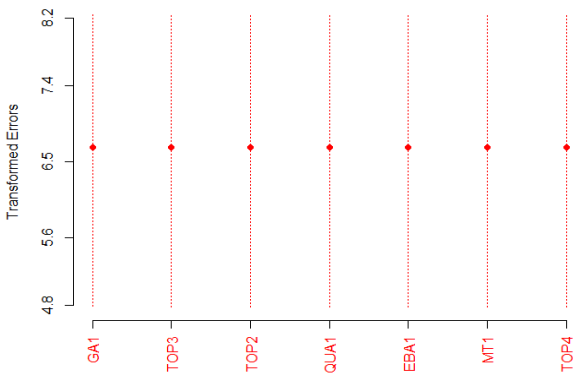
(e) Maxwell dataset

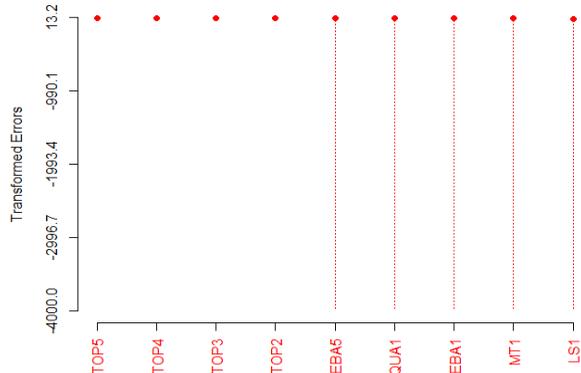
(f) China dataset

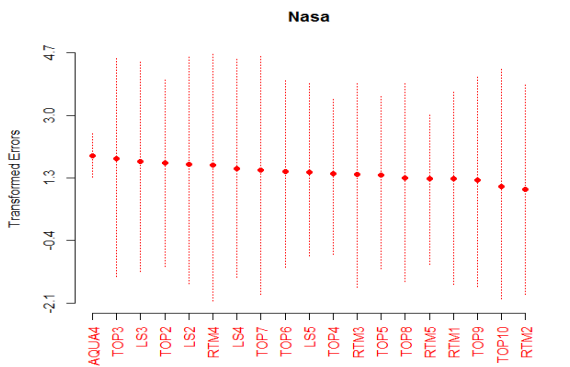
(g) Nasa dataset

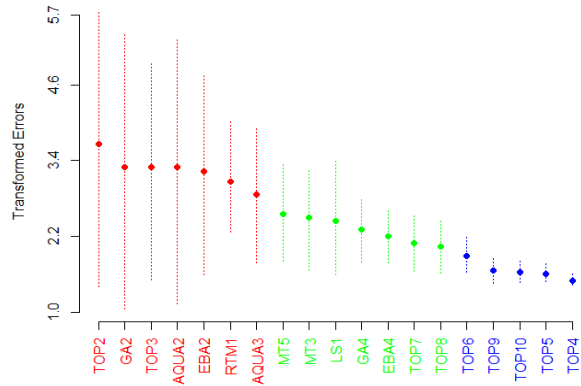
(h) Telecom dataset

Fig 4. Plot of the Scott-Knott algorithm based on Transformed absolute errors for best and ensembles

TABLE 7
The ensemble and single adjustment methods are sorted based on scores obtained by Borda counting rule. These methods have been compared over 8 datasets using 4 error measures.

| Rank | Albrecht Method | ξ | Kemerer Method | ξ | Desharnais Method | ξ | Cocomo Method | ξ | Maxwell Method | ξ | China Method | ξ | Nasa Method | ξ | Telecom Method | Ξ |
|---|---|---|---|---|---|---|---|---|---|---|---|---|---|---|---|---|
| 1 | GA5 | 1.5 | **Top3** | 1.25 | **Top5** | 1.5 | **Top5** | 1.5 | GA1 | 1.25 | LS1 | 1 | **Top9** | 1.5 | **Top10** | 1.75 |
| 2 | **Top4** | 2.25 | **Top4** | 2.5 | **Top2** | 2 | **Top3** | 2 | **Top2** | 2.5 | **Top2** | 2.5 | LS5 | 2 | **Top7** | 2.25 |
| 3 | **Top2** | 2.75 | **Top5** | 3 | LS4 | 2.5 | RTM2 | 3 | **Top3** | 3 | **Top3** | 3.25 | **Top7** | 2.5 | **Top5** | 2.5 |
| 4 | **Top5** | 4.25 | MT5 | 4 | **Top4** | 4.5 | RTM5 | 4.5 | **Top4** | 3.25 | **Top4** | 3.25 | **Top5** | 3.75 | **Top9** | 3.25 |
| 5 | **Top6** | 4.75 | **Top6** | 4.25 | **Top3** | 4.5 | **Top2** | 5.5 | MT1 | 5.25 | **Top5** | 5.5 | **Top4** | 5.5 | **Top8** | 5.25 |
| 6 | **Top3** | 5.75 | MT2 | 6.25 | MT4 | 6.5 | **Top6** | 5.5 | EBA1 | 6 | EBA5 | 6 | **Top8** | 6 | GA4 | 6.25 |
| 7 | LS2 | 7.5 | GA5 | 7.5 | GA3 | 7 | **Top4** | 6.5 | AQUA1 | 6.75 | MT1 | 7.25 | **Top6** | 7.25 | **Top6** | 7.75 |
| 8 | GA3 | 9 | EBA5 | 9 | LS1 | 7.75 | RTM4 | 8.25 | | | EBA1 | 7.5 | LS2 | 8.25 | GA1 | 7.75 |
| 9 | RTM1 | 8.75 | NN1 | 9.25 | EBA3 | 8.75 | RTM1 | 8.75 | | | AQUA1 | 8.75 | AQUA4 | 9 | RTM1 | 9 |
| 10 | AQUA5 | 9.25 | AQUA5 | 9.25 | | | LS2 | 10.25 | | | | | **Top10** | 9.25 | **Top2** | 9.75 |
| 11 | GA4 | 10.25 | **Top2** | 9.75 | | | LS3 | 10.5 | | | | | RTM5 | 11.25 | AQUA3 | 10.75 |
| 12 | | | | | | | | | | | | | **Top2** | 12 | MT3 | 12 |
| 13 | | | | | | | | | | | | | **Top3** | 13.25 | LS1 | 13 |
| 14 | | | | | | | | | | | | | RTM3 | 13.25 | EBA2 | 13.5 |
| 15 | | | | | | | | | | | | | RTM4 | 15.25 | AQUA2 | 15.25 |
| 16 | | | | | | | | | | | | | RTM1 | 15.75 | **Top4** | 16 |
| 17 | | | | | | | | | | | | | LS3 | 17.25 | **Top3** | 17 |
| 18 | | | | | | | | | | | | | RTM2 | 18.25 | EBA4 | 18.25 |
| 19 | | | | | | | | | | | | | LS4 | 18.5 | MT5 | 18.5 |

## 5 DISCUSSION

Ensemble is a machine learning method that leverages the efficiency of multi-methods to obtain better accuracy than any single method can do. The primary goal when building ensembles is the same as establishing committee of members where each method can patch mistakes done by other methods in that ensemble. In a committee, members compete amongst themselves, but at the same time, they are complementary to each other. This means that if a member's decision is not right, other members can notice and correct this decision. From the empirical analysis, it can be observed that, at least for the sample of datasets investigated, ensemble methods are good predictors with respect to Scott-Knott and Borda count. Although it must be noticed that we did not generate all possible multi-methods, the constructed ensembles occupy better positions than single methods over some datasets, specifically Kemerer, Desharnais, Cocomo, Nasa and Telecom datasets as shown in Table 7.

The procedure used in constructing ensemble was based on a definite pattern where the *TopZ* single methods were involved in all constructed ensembles. This pattern shows that there is a high probability that when one or a few methods in an ensemble make an error, the remaining methods can work together to correct this error. Further, results in Table 7 and Figure 3 would seem to reveal a number of interesting results. First, if we compare the rank of EBA variant as shown in the first column with the average rank across multiple error measures, we can notice that the rank changes of all ensemble and single methods remain stable across all error measures, over 8 datasets. Second, ensemble methods were ranked higher over all datasets as shown in Table 8, which considers several performance measures, but ranked behind according to the Scott-Knott test based on the transformed *MAE*. Table 8 shows the average ranking for ensemble methods together and single methods together over all datasets, being extracted from Table 7. It can be seen that ensemble methods have smaller average ranking, which indicates generally better overall performance across different measures.

TABLE 8
The average ranks per dataset for all ensemble and single adjustment methods according to Table 7

| | Albrecht | Kemerer | Desharnais | Cocomo | Maxwell | China | Nasa | Telecom | Average |
|---|---|---|---|---|---|---|---|---|---|
| **Ensemble methods** | 3.95 | 4.15 | 3.13 | 4.20 | 2.92 | 3.63 | 6.78 | 7.30 | 4.51 |
| **Single methods** | 7.71 | 7.34 | 6.50 | 7.54 | 4.81 | 6.10 | 12.88 | 12.43 | 8.164 |

This study has also highlighted a couple of issues: 1) number of ensemble members, and 2) number of nearest analogies used by single superior methods. The first finding that needs to be discussed is how many methods should be used to construct an ensemble method. Referring to the results we can observe that any number of methods can be used but the overall performance is determined by the ability of the aggregated methods to boost each other. Also, the methodology used in selecting single methods to form an ensemble minimizes the time and effort. One can think that we can construct all possible combinations from available methods, but however we recommend to construct ensembles from *Top2*, *Top3*,…,*TopM* best methods.

Concerning the second issue, it has been reported in previous studies that choosing the number of nearest analogies is a difficult task in using EBA. Table 9 shows the single adjustment methods and their associated best $k$ for each one (i.e. the best variant of each method as appeared first in Figure 3). Although we used a few number of analogies, we noticed that the adjustment methods behave diversely where none of them favors a particular number, so there is no definite pattern on which $k$ settings perform better. However, we can notice that GA often works better with large number of $k$, whereas RTM and MLFE work better with small $k$ value. It interesting to see that the datasets with large number of categorical features favor $k=1$ as in Maxwell dataset. However, these studies were over limited number of datasets and error measures. This question remains open and there is no clear answer, but some recent studies report that the choice of this value prior making prediction is of high importance and dependent on dataset characteristics. Moreover, each project can favor different $k$ value in comparison to other projects in the same dataset.

TABLE 9
Each adjustment methods and its best $k$ value for each dataset from Figure 3.

| | Best $k$ value | | | | | | | |
|---|---|---|---|---|---|---|---|---|
| | **Albrecht** | **Kemerer** | **Desharnais** | **Cocomo** | **Maxwell** | **China** | **Nasa** | **Telecom** |
| EBA | 5 | 5 | 3 | 5 | 1 | 1 | 5 | 4 |
| LS | 5 | 4 | 4 | 2 | 1 | 1 | 5 | 1 |
| MLFE | 3 | 2 | 3 | 3 | 1 | 1 | 2 | 1 |
| RTM | 1 | 2 | 3 | 5 | 1 | 3 | 2 | 1 |
| MT | 4 | 2 | 4 | 3 | 1 | 1 | 1 | 3 |
| AQUA | 5 | 5 | 5 | 5 | 1 | 1 | 4 | 3 |
| GA | 4 | 5 | 3 | 5 | 1 | 4 | 5 | 4 |
| NN | 5 | 1 | 5 | 3 | 1 | 5 | 1 | 3 |

Based on the above discussion, we can report from the above results that the single methods that belong to linear adjustment strategy are more accurate than those of nonlinear as confirmed in Figure 5. In Figure 5, we compare between the rankings of each EBA adjustment types using Scott-Knott test method and two-way ANOVA. The principal objective of this comparison is to investigate which type of adjustment would perform better than others. The two-way ANOVA examines the influence of two different independent factors on one dependent variable. The first independent factor of this analysis is the adjustment type (treatment) and the second factor is the number of nearest analogies ($k$) (groups). It is important to know that each treatment has five groups based on the number of nearest analogies ($k$) - for example EBA has EBA1, EBA2, EBA3 EBA4 and EBA5. So we have 8 treatments according to the number of adjustment types and 5 groups according to the value of $k$. The dependent variable is the absolute error in our analysis. Since Scott-Knott and two-way ANOVA presume provisionally that the error should be normally distributed, we should make sure that all methods' errors are transformed to be normally distributed. For that, we use one of the common transformation methods, namely the Box-Cox method. It is clear that the RTM was ranked first 3 times. Specifically, linear adjustment methods that are based on size adjustment such as LS and RTM work better than their counterparts that are based on similarity measure like AQUA. Generally, it can be remarked from Figure 5 that the non-linear adjustment method NN and MT occupied relatively worse positions over most datasets. In summary, we recommend to use ensemble methods that are constructed from linear single adjustment methods because they are frequently ranked high over nonlinear methods.

Finally, the criteria of success for any method can be seen from the practitioner's point of view as a three dimensional point: accuracy, stability and transparency. Accuracy is one of the important criteria when choosing a method (be it an ensemble method or a single method). Practitioners need methods that will produce accurate estimations – inaccurate estimations would lead to wrong decisions by the practitioners. The stability of the method across different conditions is also an important criterion to be considered. A practitioner can be more confident that a method that has been shown to be accurate and stable across different conditions will work in his/her context than a method that has shown to be unstable. Our study shows that ensembles are in general more stable than other methods. This facilitates the choice of method to be used: rather than making the difficult choice about which unstable single method to use in a given context, the practitioner can simply use a combination of several different methods (ensemble). Another criterion that a practitioner can take into account is the transparency of the model generated, i.e., how easy it is to understand the model generated by the approach. Ensemble models can lead to better accuracy and stability, as shown in the paper. However, they may be less transparent than some single methods. It would be down to the practitioner to decide whether he/she prefers a method that has been shown to achieve better accuracy and stability but is less transparent, or a method that has shown to be less accurate and less stable, but is more transparent.

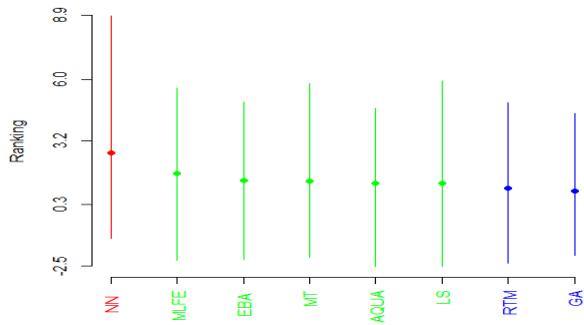
(a)　Albrecht dataset

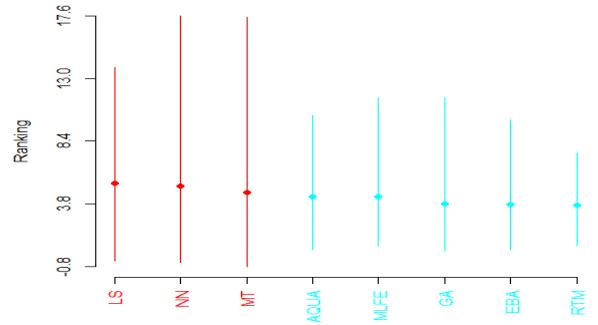
(b)　Kemerer dataset

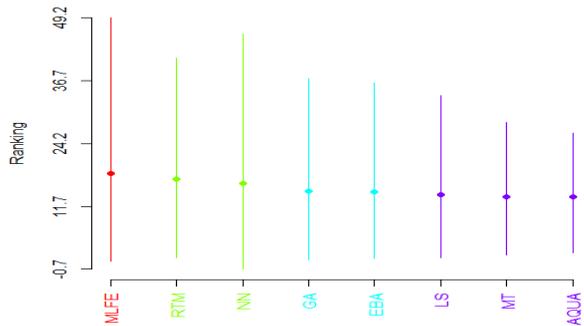
(c)　Desharnais dataset

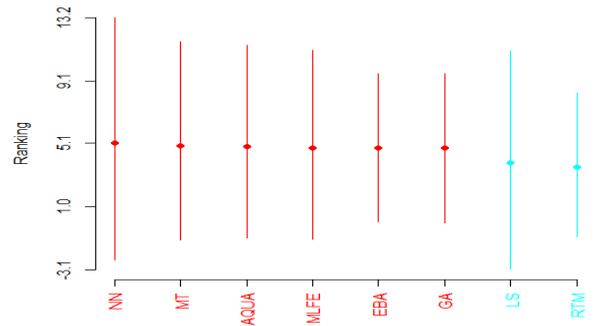
(d)　Cocomo dataset

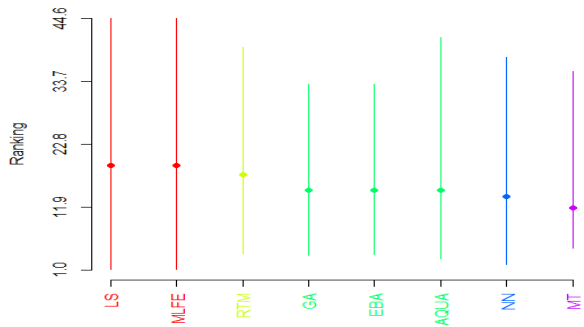
(e)　Maxwell

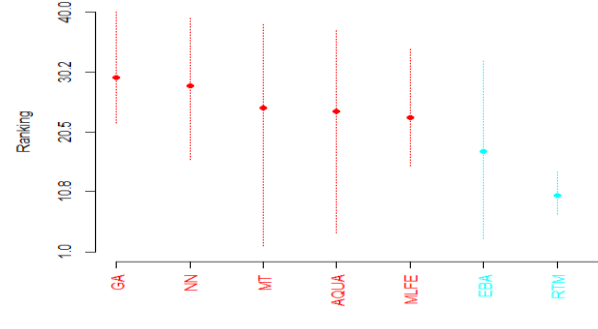
(f)　China

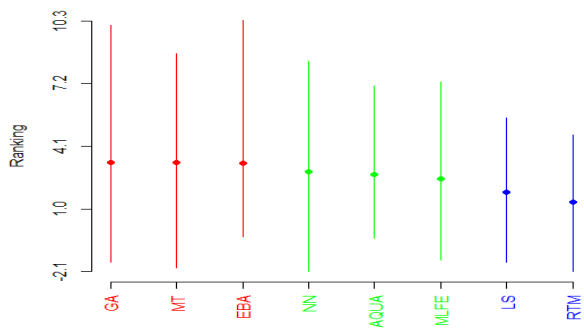
(i)　Nasa Dataset

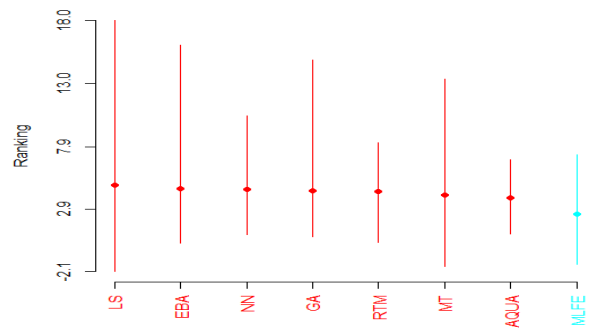
(g)　Telecom dataset

Fig 5. Plot of the Scott-Knott algorithm with two-way ANOVA for adjustment method types

# 6 THREATS TO VALIDITY

This section describes threats to validity of this research with respect to internal and external validity. The main internal validity question is: Is the variation in the dependent variable due to the changes in the independent variable? To address this issue we used 8 datasets and applied leave-one-out cross validation in our experiments so that, for each iteration, we used a different test instance and a different train set. The main advantage of the leave-one-out method is that it can be replicated with the same outcomes when using a particular dataset. On the contrary, in the cross-validation technique, a dataset is divided randomly into groups, which makes the cross-validation technique harder to be replicated. Additionally, leave-one-out validation generates higher variance estimates than cross-validation because leave-one-out performs many more tests [51]. Furthermore, since cross-validation must learn from fewer data points than the leave-one-out validation, the latter generates lower bias estimates [51]. Nonetheless, not using cross-validation in our research can be considered a threat to the validity of our results since we did not check if our findings using the leave-one-out would be the same if the cross-validation method was used instead. On the other hand, using 5 different $k$-settings for each adjustment method was reasonable since the majority of previous studies that use pre-determined number of nearest analogies recommend using $k<=5$.

Regarding the error evaluation criteria, we used several reliable error measures such as *MBRE, MIBRE, LSD*, and the *MAE* as well as the *SA* and *effect size* because *MRE, MMRE, MMER, MdMRE* were reported to be unreliable and untrustworthy. The reasons for our choice are threefold: 1) They are unbiased and practical options for the majority of researchers, and 2) using such measures enables our study to be benchmarked with previous effort estimation studies, 3) these measures as a group give us a collective decision regarding predictive performance of the used methods.

Concerning external validity, i.e. the ability to replicate our study and generalize the obtained findings, we did our best to use several datasets coming from different industrial sectors and sources, with 8 public datasets having been used in our study. The employed datasets contain a wide diversity of projects in terms of their sources, their domains and the time period they were developed in. We also believe that reproducibility of results is an important factor for external validity. Therefore, we have purposely selected publicly available datasets.

# 7 CONCLUSIONS

In this paper, we studied 8 adjustment methods existing in literature. We have conducted several experiments on 40 variants of single adjustment methods using 4 performance measures and 8 historical datasets to investigate the accuracy of ensembles on adjustment methods. Our results reveal that ensembles of adjustment methods relatively improve the prediction accuracy compared to single adjustment methods used in ABE. Therefore, we conclude that as it is always hard to identify the best single method, there will be definitely a superior multi-method constructed from strong single methods. Below we present answers to our research questions:

*RQ1. Is there evidence that ensembles improve the accuracy of adjusted EBA?* The results of our experiments show evidence that ensembles of adjustment methods can produce relatively good accuracy and occupy high ranking, but are not always superior to single methods, as confirmed in Figure 4 and Table 7. Therefore, we conclude that, as it is often hard to identify the best single method, a superior multi-method can be constructed from strong single methods. This has led us to another important question regarding number of methods that should be used in one ensemble and how to combine them. However, as it can be seen in literature, different methods would work well under different combination schema.

*RQ2. Which approach is better for adjustment, Linear of Non-Linear methods?* So far, there was no evidence suggesting the accuracy of linear methods over nonlinear methods. To provide a justifiable answer for this question, we have conducted several experiments on 40 variants of single adjustment methods using 4 error measures and 8 historical datasets. The assumption made concerning superiority suggests that the superior method should have consistent performance across all error measures, over all datasets. The results found in Figures 2 and 4 suggest that linear adjustment methods are generally better than non-linear methods. Specifically, the variants of the LSE and RTM adjustment method produce more accurate predictions than GA and AQUA varaiants, as they obtained highest and stable ranking across all experimental conditions.

*RQ3. Is there evidence that using different k analgoies makes adjustment method behave diversely?* Yes, there is evidence that changing $k$ analogies would make the adjustment method behave diversely in terms of its prediction and accuracy. This has been confirmed based on Scott-Knott test in Figure 3, where different variants of each adjustment method would belong to different clusters. Concerning the $k$ nearest analogies, there is no clear answer about choosing that optimal number, but our findings suggest that $k=5$ would seem to work better for GA and $k=1$ for all methods over Maxwell as shown in Table 9. In the ideal case, different adjustment methods could perform optimally under the variability of $k$ values for every single project. In other words, it would be better to use a technique that can catch the preference of each project and be able to produce better predictive accuracy. The results obtained do not support those implications because we used a fixed number of $k$ values for all projects in a dataset used in particular experiments.

## ACKNOWLEDGMENT

Mohammad Azzeh and Ali Bou Nassif are grateful to the Applied Science University, Amman, Jordan, for the financial support granted to carry out this research. Leandro Minku is grateful to EPSRC for the financial support given through the Grant No. EP/J017515/1.